\documentclass[aps,prd,preprint,amsmath,amssymb]{revtex4}

\usepackage{graphicx}
\usepackage{dcolumn}
\usepackage{bm}

\begin{document}







\def\beq{\begin{equation}}
\def\eeq{\end{equation}}
\def\bea{\begin{eqnarray}}
\def\eea{\end{eqnarray}}
\def\ben{\begin{enumerate}}
\def\een{\end{enumerate}}
\def\la{\langle}
\def\ra{\rangle}
\def\a{\alpha}
\def\b{\beta}
\def\g{\gamma}\def\G{\Gamma}
\def\d{\delta}
\def\e{\epsilon}
\def\phi{\varphi}
\def\k{\kappa}
\def\l{\lambda}
\def\m{\mu}
\def\n{\nu}
\def\o{\omega}
\def\p{\pi}
\def\r{\rho}
\def\s{\sigma}
\def\t{\tau}
\def\L{{\cal L}}
\def\S{\Sigma }
\def\gsim{\; \raisebox{-.8ex}{$\stackrel{\textstyle >}{\sim}$}\;}
\def\lsim{\; \raisebox{-.8ex}{$\stackrel{\textstyle <}{\sim}$}\;}
\def\gtrsim{\gsim}
\def\lessim{\lsim}
\def\loc{{\rm local}}
\def\vm{v_{\rm max}}
\def\bh{\bar{h}}
\def\del{\partial}
\def\nab{\nabla}
\def\half{{\textstyle{\frac{1}{2}}}}
\def\fourth{{\textstyle{\frac{1}{4}}}}

\title{Spherical Solutions in Einstein-Aether Theory:
Static Aether and Stars}

\author{Christopher Eling}
 \email{cteling@physics.umd.edu}
 \author{Ted Jacobson}
\email{jacobson@umd.edu}
 \affiliation{Department of Physics, University of Maryland\\ College Park, MD 20742-4111 USA}


\begin{abstract}

The time independent spherically symmetric solutions of General
Relativity (GR) coupled to a dynamical unit timelike vector are
studied. We find there is a three-parameter family of solutions with
this symmetry. Imposing asymptotic flatness restricts to two
parameters, and requiring that the aether be aligned with the
timelike Killing field further restricts to one parameter, the total
mass. These ``static aether" solutions are given analytically up to
solution of a transcendental equation. The positive mass solutions
have spatial geometry with a minimal area 2-sphere, inside of which
the area diverges at a curvature singularity occurring at an
extremal Killing horizon that lies at a finite affine parameter
along a radial null geodesic. Regular perfect fluid star solutions
are shown to exist with static aether exteriors, and the range of
stability for constant density stars is identified.

\end{abstract}

\maketitle

\section{Introduction}
\label{intro}

For nearly a century Lorentz invariance has been a basic
assumption in theoretical physics. However, various approaches to
the problem of quantum gravity question whether Lorentz symmetry
is truly fundamental. In this context, it is useful to consider
low energy Lorentz violating (LV) models from the point of view of
effective field theory. It makes sense to distinguish LV effects
in the matter sector from LV effects in the much more weakly
coupled (at low energies) gravitational sector. LV effects in the
matter sector are tightly constrained by high precision tests of
Lorentz invariance \cite{Mattingly:2005re}. Lorentz violation in
the gravitational sector was  studied by Will and
Nordvedt~\cite{LVmetr} in the early 1970's, and it was pursued in
the 1980's by Gasperini~\cite{Gasp} motivated by singularity
prevention, and by Kostelecky and Samuel~\cite{Kostelecky:1989jw}
motivated by the possibility of spontaneous LV in string theory.
More recently a number of approaches to incorporating LV into
gravity have been explored; see for example
Refs.~\cite{Clayton:1999zs,  Jacobson:2000xp, Arkani-Hamed:2003uy,
Gripaios:2004ms,Bluhm:2004ep, Heinicke:2005bp,Rubakov:2006pn} and
references therein.

Here we study the theory of a unit timelike vector field $u^a$
coupled only to gravity, as an LV modification of General Relativity
(GR). To preserve general covariance this field must be dynamical.
It can be thought of as the 4-velocity of a preferred frame, which
breaks boost symmetry at every point of spacetime. This property it
has in common with the 19th century concept of the aether, and like
that aether it is also a dynamical system. Hence a suitable name for
it is the ``aether", although it has nothing to do with
electromagnetism. Since the aether is coupled to Einstein GR we
refer to the theory as ``Einstein-Aether" theory; we sometimes use
the nickname ``ae-theory" for short.  For a review of the
motivation,  history and recent status of the Einstein-Aether
theory, see~\cite{Eling:2004dk} and the references therein.

Observational constraints on the Einstein-Aether theory have been
determined from PPN analysis~\cite{Eling:2003rd,
Graesser:2005bg,Foster:2005dk}, stability and energy
positivity~\cite{Jacobson:2004ts,Lim:2004js, Elliott:2005va,
Eling:2005zq}, primordial nucleosynthesis~\cite{Carroll:2004ai},
and Cerenkov radiation~\cite{Elliott:2005va}. The combined
constraints from all of these are reviewed in
Ref.~\cite{Foster:2005dk}, and constraints from radiation damping
are discussed in~\cite{Foster:2006az}. Also important for
determining theoretical viability and observational constraints
are the properties the spherically symmetric static solutions.
These solutions were previously obtained in the the special case
where the aether dynamics is
Maxwell-like~\cite{Kostelecky:1989jw,Jacobson:2000xp}. It was
shown in~\cite{Jacobson:2000xp} that the Reissner-Nordstrom metric
in a spherically symmetric static gauge with fixed norm is a
solution, and it was claimed (incorrectly, as shown here) that
this is the only solution in that special case. The asymptotic
weak field limit of the general case was studied in
Ref.~\cite{Eling:2003rd}, where it was found that there is a
two-parameter family of asymptotically flat spherical, static
solutions. A thorough examination of the fully nonlinear solutions
has not been carried out before.

In this paper and its companion (devoted to black
holes)~\cite{Eling:2006ec} we complete a general survey of the
time-independent spherically symmetric solutions. In section
\ref{sec:action} of this paper the action and field equations for
the Einstein-Aether theory are reviewed. Specializing to
stationary, spherical symmetry, section \ref{Analysis} studies the
structure of the field equations as ordinary differential
equations and shows that locally there is a three parameter family
of vacuum solutions. Imposing asymptotic flatness reduces the
number of parameters to two.

In section \ref{static} the general solution in which the aether
vector is aligned with the timelike Killing field is found. Being a
unit vector, the aether is completely determined by the metric in
this case. This solution is asymptotically flat and described by one
free total mass parameter. For negative total mass there is a naked
singularity at the origin. The positive mass solution has a
wormhole-like spatial geometry, reaching a minimum area 2-sphere at
some radius like the Schwarzschild solution. Unlike in the
Schwarzschild solution this throat is not on a horizon. Inside the
throat the spheres re-expand to infinitely large size in finite
affine parameter along a radial null geodesic and finite or infinite
proper distance depending on the coupling parameters in the
Lagrangian. When the distance is finite the internal infinity is
singular, and it occurs at a would-be extremal Killing horizon. When
the distance is infinite the metric is asymptotically singular.

In section \ref{stars} it is first shown that pure aether stars do
not exist, i.e. there are no asymptotically flat self-gravitating
aether solitons with a regular origin. (Given the results of the
previous section, regular aether stars could only possibly have
existed if the aether were not everywhere aligned with the Killing
vector.) Next it is shown that in the presence of a perfect fluid,
regular asymptotically flat star solutions exist and are
parameterized (for a given equation of state) by the central
pressure. For the case of constant density the star solutions are
found by matching numerical integration for the interior to the
vacuum solution found in section \ref{static}. As in GR, for a
given density there is a maximum mass. Utilizing the critical
behavior of the mass as a function of stellar radius $R$ it is
shown that if they are stable at small mass, these stars are
unstable beyond the maximum mass. We conclude in section
\ref{discussion} with a discussion of open questions raised by
these results.

\section{Einstein-Aether action}
\label{sec:action}

The action for Einstein-Aether theory is
the most general diffeomorphism invariant
functional of the spacetime metric $g_{ab}$ and aether field $u^a$
involving no more than two derivatives,
\beq S = \frac{1}{16\pi G}\int \sqrt{-g}~ L ~d^{4}x
\label{action} \eeq
where
\beq L = -R-K^{ab}{}_{mn} \nabla_a u^m \nabla_b u^n\\ -
\lambda(g_{ab}u^a u^b - 1). \eeq
Here $R$ is the Ricci scalar,  $K^{ab}{}_{mn}$ is defined as
\beq K^{ab}{}_{mn} = c_1
g^{ab}g_{mn}+c_2\delta^{a}_{m}\delta^{b}_{n}
+c_3\delta^{a}_{n}\delta^{b}_{m} +c_4u^au^bg_{mn} \eeq
where the $c_i$ are dimensionless constants, and $\lambda$ is a
Lagrange multiplier enforcing the unit timelike constraint. This
constraint restricts variations of the aether to be spacelike, hence
ghosts need not arise. A term of the form $R_{ab} u^a u^b$ is not
explicitly included as it is proportional to the difference of the
$c_2$ and $c_3$ terms in (\ref{action}) via integration by parts.
The metric signature is $({+}{-}{-}{-})$ and the units are chosen so
that the speed of light defined by the metric $g_{ab}$ is unity. In
spherical symmetry the aether is hypersurface orthogonal, hence it
has vanishing twist $\omega_a = \epsilon_{abcd} u^b \nabla^c u^d$.
When $u^a$ is a unit vector the square of the twist is a combination
of the $c_1$, $c_3$, and $c_4$ terms in the action ({\ref{action}),
\beq \omega_a \omega^a = -(\nabla_a u_b)(\nabla^a u^b) + (\nabla_a
u_b)(\nabla^b u^a)+ (u^b \nabla_b u_a)(u^c \nabla_c u^a).\eeq
The $c_4$ term can thus be absorbed by making the replacements $c_1
\rightarrow c_1 + c_4$ and $c_3 \rightarrow c_3 - c_4$.

Following the observational constraints we assume here when studying
the fluid star solutions that the only significant coupling of $u^a$
to matter is through a universal ``matter metric" $g^{\rm
matter}_{ab}=g_{ab}+\sigma u_a u_b$, where $\sigma$ is a constant.
Replacing $g_{ab}$ by $g^{\rm matter}_{ab}$ as the independent
metric field in the action returns an action with the same form as
(\ref{action}) but with new values of the constants $c_{1,2,3,4}$
that depend on $\sigma$~\cite{Foster:2005ec}. Hereafter we assume
that such a field redefinition has already been performed, so that
$g_{ab}$ is the metric to which matter couples universally. The
absence of any other coupling of $u^a$ to matter has no theoretical
justification in this purely phenomenological approach, and may be
regarded as unnatural. However our goal here is just to explore
consequences of gravitational Lorentz violation in a
phenomenologically viable setting. It remains an open question
whether this can emerge as an approximation to a more fundamental
underlying theory.

The field equations from varying (\ref{action}) plus a matter action
(coupled only to the metric) with respect to $g^{ab}$, $u^a$ and
$\lambda$ are given by
\begin{eqnarray}
G_{ab} &=& T^{(u)}_{ab}+8\pi G T^{M}{}_{ab}\label{AEE}\\
\nab_a J^{a}{}_m-c_4 \dot{u}_a \nab_m u^a &=& \l u_m,
\label{ueqn}\\
g_{ab} u^a u^b &=& 1, \label{constraint}
\end{eqnarray}
where
\beq J^a{}_{m} = K^{ab}{}_{mn} \nabla_b u^n. \label{Jdef}\eeq
The aether stress tensor is given by~\cite{Eling:2003rd}
\bea T^{(u)}{}_{ab}&=&\nab_m(J_{(a}{}^m u_{b)}-J^m{}_{(a} u_{b)}- J_{(ab)}u^m) \nonumber\\ &&
+ c_1\, \left[(\nab_m u_a)(\nab^m u_b)-(\nab_a u_m)(\nab_b
u^m) \right]\nonumber\\ &&+ c_4\, \dot{u}_a\dot{u}_b\nonumber\\
&&+\left[u_n(\nab_m J^{mn})-c_4\dot{u}^2\right]u_a u_b \nonumber\\
&&-\frac{1}{2} L_u g_{ab}, \label{aetherT}\eea
where $L_{u} = -K^{ab}{}_{mn} \nabla_a u^m \nabla_b u^n$. The
Lagrange multiplier $\lambda$ has been eliminated from
(\ref{aetherT}) by solving for it via the contraction of the aether
field equation (\ref{ueqn}) with $u^a$.

Some words about terminology are in order. Spacetimes admitting a
timelike Killing vector field $\xi^a$ are generally called {\it
stationary}. In the special case where $\xi^a$ is hypersurface
orthogonal, and therefore invariant under a time reflection $t
\rightarrow -t$, the spacetime is said to be \textit{static}. A
stationary aether field $u^a$ on a stationary spacetime is one whose
Lie derivative with respect to $\xi^a$ vanishes. If the spacetime is
static, one might be tempted to say the aether is ``static", however
this is not really appropriate since the aether itself breaks the
Killing time reflection symmetry. The solutions studied in this
paper involve a static metric coupled to a stationary aether. This
general situation will be called here ``stationary spherical
symmetry". An important special case occurs when the aether is
parallel to the Killing vector. We refer to this special case as
``static aether". Such an aether changes sign under the Killing time
reflection, however the action (\ref{action}) is invariant under
$u^a\rightarrow -u^a$ so the sign of $u^a$ has no physical meaning.
Note that regular black holes cannot have static aether fields since
the Killing vector is null, not timelike on the horizon.

\section{Classification of stationary spherical solutions}
\label{Analysis}

Stationary spherically symmetric solutions describe, for example, a
black hole or the exterior of a time-independent star. In spherical
symmetry all stationary metrics are static~\cite{WaldBook}. The line
element can be written in Schwarzschild type coordinates,
\beq ds^2 = e^{A(r)} dt^2 - B(r) dr^2 - r^2 d\Omega^2, \label{Schw}
\eeq
and the aether field takes the form
\beq u = a(r)\partial_t + b(r)\partial_r. \label{uform}\eeq
The unit constraint on $u^a$ becomes
\beq e^{A(r)}a(r)^2-B(r)b(r)^2 = 1, \label{constr1} \eeq
which can be used for example to eliminate $b(r)$. The $t$-component
of the aether field equation (\ref{ueqn}) can be used to solve for
$\l$ in this case, and the remaining field equations reduce to five
ODE's: the $tt$, $rr$, $tr$, and $\theta\theta$ components of the
metric field equation and the $r$ component of the aether field
equation. These five equations involve the eight functions $\{A'',
A', A, B', B, a'', a', a\}$, where prime denotes differentiation
with respect to the argument $r$ which is suppressed. The equations
are too complicated to be worth writing down here, so we shall just
describe their structure. Using the $tt$ and $\theta\theta$ metric
equations along with the $r$ component of the aether equation, one
can solve for $A''$, $a''$, and $B'$ in terms of the remaining five
functions $\{A', A, B, a',a\}$. It turns out that only one
additional piece of information remains in the $tr$ and $rr$
equations, which can be used to solve (for example) for $B$ in terms
of $\{A', A, a',a\}$. Finally, $A(r_0)$ at any given value $r=r_0$
can be chosen at will by allowing for an appropriate scaling of the
$t$ coordinate. At a given $r_0$ value, the remaining three values
$\{A'(r_0), a'(r_0),a(r_0)\}$ then determine a (local) solution by
integration with respect to $r$. This shows that there is in general
a three-parameter family of spherically symmetric stationary
solutions.

To illustrate this reasoning in a more familiar setting, we can
apply it to the field equations of pure GR in Schwarzschild
coordinates,
\begin{eqnarray}
G_{tt} &\propto& r B'-B+B^2 = 0 \\
G_{rr} &\propto& r A'-B+1 = 0 \\
G_{\theta\theta} &\propto& 2 r A''B+ A'(2B-rB') +r A'^2 B - 2 B' =
0.
\end{eqnarray}
These can be used to solve for $B'$, $B$, and $A''$ in terms of $A'$
and $A$. Using the freedom to scale $t$ the initial value $A(r_0)$
can be fixed at will, so we recover the well-known fact that static
spherically symmetric solutions in GR are characterized by one free
parameter, in this case the value of $A'(r_0)$. In the
Einstein-Aether theory the aether vector and its derivative provide
two additional degrees of freedom at each point.

Birkhoff's theorem in GR states that the only spherically symmetric
solution is static and given (up to coordinate freedom) by the
Schwarzschild metric. The radial tilt of the aether provides another
local degree of freedom in ae-theory, so spherical solutions need
not be time-independent. But, as we have seen, even restricting to
stationary spherically symmetry ae-theory has more solutions. In
this paper we will focus primarily on the static aether solutions,
which form a one-parameter family. Black hole solutions, which
comprise a different family, are studied in a companion paper.

Numerical integration of the ae-theory field equations as ODE's out
from some arbitrary point $r_0$ with generic initial conditions
yields singularities in $A(r)$, $B(r)$, and $a(r)$. However, there
is a two-parameter family of asymptotically flat solutions. This was
first found in Ref.~\cite{Eling:2003rd} using a perturbative
expansion about infinity. Asymptotic flatness was imposed there by
assuming regular power series expansions about $x=1/\rho=0$, where
$\rho$ is the isotropic radial coordinate. Asymptotic flatness can
also be imposed using the ``shooting method". This is simple to
implement here since it is only necessary to tune one of the three
initial values $\{ A'(r_0),a'(r_0), a(r_0)\}$ so that, for example,
$A(r)$ approaches a constant value as $r\rightarrow\infty$. The
field equations then automatically enforce the remaining asymptotic
flatness conditions. In GR, by contrast, asymptotic flatness is a
consequence of the vacuum field equations without any tuning of
initial data, so the one-parameter family of local (Schwarzschild)
solutions is automatically asymptotically flat.

\section{Static Aether}
\label{static}

In this section we obtain the static aether solution, where the
aether vector $u^a$ is proportional to the timelike Killing field
$\xi^a$ and therefore entirely determined by the metric.

\subsection{Field equations with static aether}
Using the
Schwarzschild type coordinates in (\ref{Schw}) and (\ref{uform}),
the static aether has $b(r)=0$ and $a(r)= \exp(-A(r)/2)$, i.e.
\beq u = e^{-A/2}\partial_t. \label{statuform} \eeq
In this case the $c_2$ and $c_3$ terms drop out of the field
equations. To see why, note that  (\ref{statuform}) implies
$\nabla_a u^a=0$, so all variations of $c_2$ term in the action
(\ref{action}) vanish. In addition, the normalization $u_a u^a=1$
implies $u_b\nabla_a u^b=0$. These conditions together with
spherical symmetry imply that the derivative of $u^b$ has the form
\beq
\nabla_a u^b= u_a s^b,
\label{staticdu}
\eeq
where $s^b$ is a radial vector orthogonal to $u^a$. (We note in
passing that contraction of (\ref{staticdu}) with $u^a$ reveals that
$s^b$ is the acceleration of the aether worldlines.) Therefore $
(\nabla_a u^b)(\nabla_b u^a)$ vanishes, so the variation of the
volume element in the $c_3$ term of the action (\ref{action})
vanishes. The remaining variation of the $c_3$ term is proportional
to $u_a s^b \d(\nabla_b u^a)= s^b\d(u_a\nabla_b u^a)$, which
vanishes for all variations $(\d g_{ab},\d u^a)$ preserving the
normalization $g_{ab} u^a u^b = 1$. Moreover, as explained in
Section~\ref{sec:action}, in spherical symmetry the $c_4$ term can
be absorbed into the $c_1$ and $c_3$ terms, hence the solutions with
static aether are fully characterized by the case with only $c_1$
non-zero.

When only $c_1$ is nonzero the aether field equation (\ref{ueqn})
reduces to
\beq
c_1\nabla^a\nabla_a u^b = \l u^b.
\eeq
Using (\ref{staticdu}) this becomes
\beq c_1 u^a\nabla_a s^b = \l u^b. \label{aetherstats} \eeq
Contraction of the left hand side of (\ref{aetherstats})
with $s_b$ is proportional to $u^a\nabla_a s^2$,
which vanishes since $s^2$ is a scalar that must be constant
along the Killing direction parallel to $u^a$. Therefore
both sides are parallel to $u^b$, so the aether equation
only determines $\l$. Contracting both sides
of (\ref{aetherstats}) with $u_b$ we find
\beq
\l=-c_1 s^2,
 \eeq
having made use of (\ref{staticdu}) and
$u^b\nabla_a s_b = -s^b\nabla_a u_b$, which follows
from $u^b s_b=0$.

The metric field equation is $E_{ab} = G_{ab}-T^{u}_{ab}=0$, and the
$tt$, $rr$, and $\theta\theta$ components of $E_{ab}$ are given by
\begin{eqnarray}
E_{tt}&=&(e^A/r^2B)\Bigl[(-1+B+rB'/B)
-\n(8 r A' + r^2 A'^2 -2r^2 A' B'/B+4 r^2 A'')\Bigr]\\
E_{rr}& =&r^{-2}(1-B+ rA' +\n r^2 A'^2)\\
E_{\theta\theta}&=&B^{-1}\Bigl[ (2rA'-2rB'/B+ r^2 A'^2 -r^2 A'B'/B
+2r^2A'')/4-\n r^2 A'^2\Bigr],
\end{eqnarray}
where for notational convenience we have introduced the symbol
\beq \n=\frac{c_1}{8}. \eeq
Using the $E_{rr}$ equation one can solve for $B$,
\beq B = 1+rA'+ \n\,  r^2A'^2, \label{Beqn}\eeq
Substituting this solution for $B$ into the
$E_{tt}$ and $E_{rr}$ equations, we find that the
equations are redundant and the system is described by the
second order ODE
\beq r^2A'' +  2rA' + r^2 A'^2 +   \n\,  r^3 A'^3=0.
\label{staticeom}\eeq
A constant shift of $A$ can be absorbed by a scaling of the $t$
coordinate, hence there is just a one parameter family of solutions.
As in GR, the solutions in this family are all asymptotically flat.

\subsection{Static aether solutions: general analysis}
To solve (\ref{staticeom}) we define the function $Y(r)$
by
\beq
Y=rA',
\label{A'}
\eeq
in terms of which the solution for $B$ becomes
\beq
B=1+Y+\n Y^2
\label{B}
\eeq
and Eqn. (\ref{staticeom}) for $A$ becomes
\beq
dY/dr=-(Y/r)(1+Y+\n Y^2)
\label{Y'}
\eeq
The problem is thus reduced to quadratures: integration
of this equation yields $Y(r)$, which also directly yields
$B$ via (\ref{B}). To determine $A$ we combine (\ref{A'}) and (\ref{Y'})
to obtain
\beq
dA/dY = -1/(1+Y+\n Y^2),
\label{dA/dY}
\eeq
which yields $A(Y)$ by integration.

The character of the solutions is evidently controlled by the
roots of $B$,
\beq
Y_\pm=(-1\pm\sqrt{1-4\n})/(2\n),
\eeq
in terms of which we have \beq B=\n(Y-Y_-)(Y-Y_+). \label{Bfactors}
\eeq The nature of the roots depends on the value of $\n$. We
consider here only positive $\n$, since that is required by
positivity of the energy of linearized spin-0
waves~\cite{Eling:2005zq}, and we restrict to $\n<1/4$ since the
Newton constant $G_{\rm N}=G/(1-c_{14}/2)$ becomes negative beyond
this limit. In the pure $c_1$ case also the
stability~\cite{Jacobson:2004ts}
 or positive energy~\cite{Eling:2005zq}
of linearized waves requires $c_1<1$, or
$\n<1/8$.
One can visualize the roots graphically:
they are the intersections of
the line $Y+1$ with the inverted parabola $-\n Y^2$.
When $\n=1/4$
the parabola is tangent to the line, and the two
roots coincide at $Y=-2$. The larger root approaches
$-1$ as $\n\rightarrow0$,
while the smaller root approaches $-\infty$, hence
in the range $0\le\n<1/4$ the roots fall within
the ranges
\beq
-\infty\le Y_-<-2, \qquad -2<Y_+\le-1.
\label{rootranges}
\eeq
Note that $Y_-=1/(\n Y_+)$, and $\n=-(1+Y_+)/Y_+^2$.

We can integrate (\ref{dA/dY}) and (\ref{Y'})
to find both $A$ and $Y$ using
the factorization (\ref{Bfactors}) and
partial fractions. The result is
\beq
N=e^A=\left(\frac{1-Y/Y_-}{1-Y/Y_+}\right)^{\frac{-Y_+}{2+Y_+}}
\label{N}
\eeq
and
\beq
\frac{r_{\rm min}}{r}=\left(\frac{Y}{Y-Y_-}\right)\left(\frac{Y-Y_-}{Y-Y_+}\right)^{\frac{1}{2+Y_+}},
\label{C/r}
\eeq
where $r_{\rm min}$ is an integration constant.
The graph of $r/r_{\rm min}$ vs. $Y$ is
plotted in Fig.~\ref{fig:rmin/r},
for the case $c_1=1$.
The values of the sphere radius $r$ and
metric functions $B$ and $N$
at the special values of $Y$ are
given in Table~\ref{criticalY's}.
\begin{figure}[htb]
\includegraphics[width=7in]{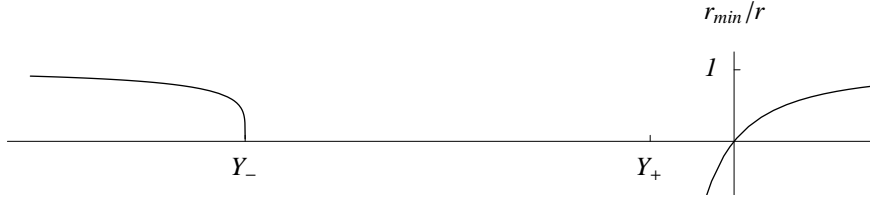}
\caption{\label{fig:rmin/r} Graph of $r/r_{\rm min}$ vs. $Y$
for $c_1=1$. The curves approach 1 asymptotically on both sides.
The range $(Y_+,0)$ defines a negative mass solution with naked singularity at
$Y=Y_+$ and asymptotically flat region at $Y\rightarrow0$.
The range $(0,\infty)$ defines a positive mass solution, with a minimal
2-sphere as $Y\rightarrow\infty$. The range $(-\infty,Y_-)$ continues that
solution to the other side of the minimal sphere, with a singularity at
a sphere of infinite radius at $Y=Y_-$. This sphere lies at finite radial distance
if $c_1<3/2$. There is no solution with timelike aether
in the range $(Y_-,Y_+)$ since the radial coordinate
is timelike there (see section \ref{nosoln}).}
\end{figure}

\begin{table}[ht]
 \caption{\label{criticalY's}Sphere radius and metric functions
at special $Y$ values.}
\begin{center}
\begin{tabular}{c||c|c|c|c}
$Y$~~~&~~~$Y_-$~~~&~~~$Y_+~~~$&~~~~0~~~~&~~~$\pm\infty$~~~\\
\hline
$r$~~~&$\infty$&0&$\infty$&$r_{\rm min}$\\
$B$~~~&0&0&1&$\infty$\\
$N$~~~&0&$\infty$&1&$>0$\\
\end{tabular}
\end{center}
\end{table}

\subsubsection{The GR limit: Schwarzschild solution}
To help to interpret the general case,
we consider first the pure GR limit $c_1=0$,
for which $Y_+=-1$ and $Y_-=-\infty$. The solution is then
\bea
B&=&1+Y\\
N&=&1/(1+Y)\\
r_{\rm min}/r&=&Y/(1+Y).
\eea
This is just the Schwarzschild solution, with $Y=1/(r-r_{\rm
min})$ and $r_{\rm min}=2M$. Spatial infinity corresponds to
$Y=0$, and as $Y\rightarrow\infty$  the radius decreases to
$r_{\rm min}$ at the bifurcation surface of the horizon. The other
side of the wormhole is here labelled by the same values of $Y$.
The range $-\infty<Y<-1$ corresponds to the future wedge of the
black hole interior, where the Killing vector is spacelike. The
remaining range $-1<Y<0$ is also significant. It corresponds to
the negative mass Schwarzschild solution.

\subsubsection{Static aether solutions for generic $c_1$}
\label{genericc1}
For generic values of $c_1$ the limit $Y\rightarrow0$
still corresponds to an asymptotically flat spatial infinity,
where the limiting form of the solution is
\bea
B&=&1+Y +\cdots\\
N&=&1-Y+\cdots\\
Y&=&2M/r+\cdots,
\eea
and the mass $M$ is related to the minimum
radius by
\beq
r_{\rm min}/2M=(-Y_+)^{-1}(-1-Y_+)^{(1+Y_+)/(2+Y_+)}.
\label{M}
\eeq
This ratio grows smoothly from 1 for $c_1=0$, to about $1.23$ for
$c_1=1$, and reaches $e/2\approx 1.4$ for $c_1=2$.

Series solution in powers of $x=2M/r$ yields
\bea
B&=&1+x + (1+\n)x^2 +\cdots\\
N&=&1-x-(\n/6)x^3+\cdots\\
Y&=&x+x^2 +(1+\n/2)x^3+\cdots.
\eea
To leading order in $1/r$ this agrees with the
Schwarzschild solution, as already seen
previously in~\cite{Eling:2003rd}, where
the two Eddington-Robertson-Schiff PPN
parameters $\gamma$ and
$\beta$ were found to agree with the GR value of 1.
This completes our characterization of the asymptotically
flat region. What happens when we follow the solution
to smaller values of $r$?

The answer depends on the range of $Y$
considered. For $Y\in(Y_+,0)$,
equation (\ref{C/r}) or its graph in Fig.~\ref{fig:rmin/r}
indicate that $r_{\rm min}$
must be {\it negative}, which according
to (\ref{M}) implies a negative total mass $M<0$.
In this case there is a naked singularity at
$Y=Y_+$ ($r=0$) connected to
an asymptotically flat region at
$Y=0$ ($r=\infty$), like in the negative mass Schwarzschild
solution.

For positive $Y$ the solution is different.
It is seen again from (\ref{Y'}) or its graph that $Y$ grows
monotonically as $r$ decreases.
As $Y\rightarrow\infty$ the r.h.s. of (\ref{C/r}) goes to $1$,
so this limit for $Y$ corresponds to a minimum radius $r_{\rm min}$,
just as in the case of the Schwarzschild solution. However,
the solution behaves quite differently from Schwarzschild.
First, instead of $N(r_{\rm min})=0$ we have
\beq
N(r_{\rm min})=\left(\frac{Y_+}{Y_-}\right)^{-Y_+/(2+Y_+)},
\eeq
so the minimal 2-sphere does not sit at a Killing horizon. The value
of $N(r_{\rm min})$ grows smoothly from $0$ for $c_1=0$ to about
$0.083$ for $c_1=1$ and reaches $e^{-2}\approx0.135$ for $c_1=2$.
(Recall that in the GR limit we have $Y_-\rightarrow-\infty$.)
Another difference due to the finiteness of $Y_-$ is that the
solution continues with negative $Y$ values, with the two values
$Y=\pm\infty$ identified. According to (\ref{Y'}), as $Y$ grows from
$-\infty$ up to $Y_-$, $r$ increases from $r_{\rm min}$ to $\infty$.
Therefore the ``interior" of the minimal 2-sphere flares out to
infinite radius as in the Schwarzschild solution. But unlike the
Schwarzschild case, now two values of $Y$ correspond to each $r$,
and the ``interior" geometry is not equivalent to the exterior. In
fact the difference is quite dramatic: at the internal infinity both
$N$ and $B$ go to zero, whereas they both approach one in the
asymptotically flat region.

The Carter-Penrose diagram for this solution is the square diamond
in Fig.~\ref{Carter} with asymptotically flat past and future null
infinity on the lower and upper edges bounding the right hand side
and a singularity on both edges $S^\pm$ and at $S^0$ bounding the
left hand side. $S^\pm$ are singular Killing horizons with vanishing
surface gravity. The proper distance to $S^0$ along a constant $t$
surface is finite if $0<c_1<3/2$ and infinite if $3/2<c_1<2$, while
for any $c_1$ the affine parameter to $S^\pm$ along a radial light
ray is finite, as we now demonstrate.
\begin{figure}
 \includegraphics[angle=0,width=10cm]{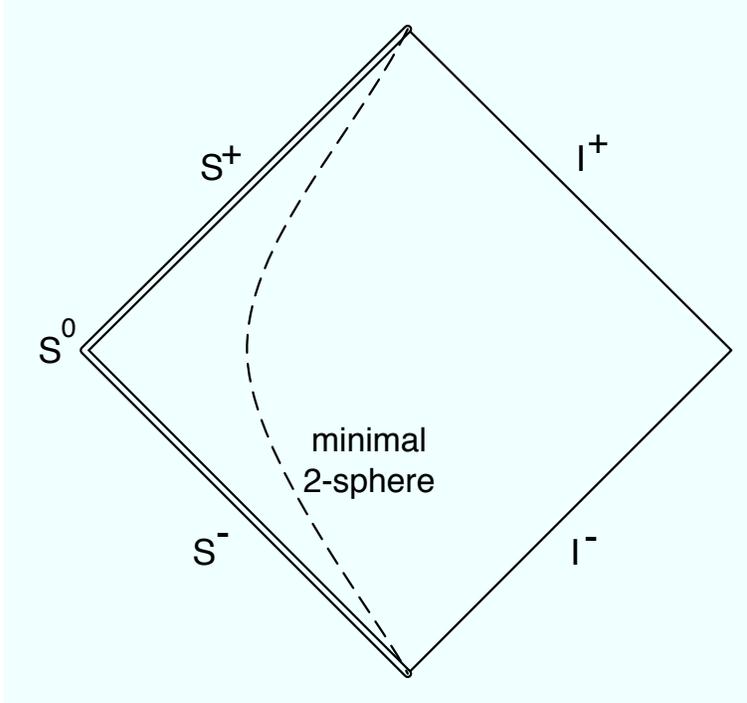}\\
\caption{\label{Carter}Carter-Penrose diagram of the static aether
solution. The left hand edge corresponds to spheres of infinite
radius and is singular.}
\end{figure}

As the internal infinity at $Y_-$ is approached, the proper radial
distance on a constant $t$ surface behaves as
\beq \frac{dl}{dr}=B^{1/2}\sim(Y_- -Y)^{1/2}. \eeq
On the other hand (\ref{C/r}) shows that in this limit
the relation between $Y$ and $r$ is
\beq (Y_- - Y)\sim r^{\frac{2+Y_+}{1+Y_+}}, \eeq
so we have
\beq
\frac{dl}{dr}=r^{\frac{2+Y_+}{2+2Y_+}}.
\label{dl/dr}\eeq
The exponent of $r$ in (\ref{dl/dr}) is always negative, and it is
equal to $-1$ when $Y_+=-4/3$, which corresponds to $\n=3/16$, i.e.
$c_1=3/2$. For $c_1<3/2$ the radial distance to $S^0$ is finite and
there is a curvature singularity at $r=\infty$ that shows up, for
example, in the square of the Riemann tensor.

Along a radial null geodesic the quantity $N\dot{t}=N dt/d\l$ is
conserved, where $\l$ is an affine parameter. Together with the
lightlike condition $N\dot{t}^2-B\dot{r}^2=0$ this implies that as
the internal infinity is approached the affine parameter behaves as
\beq d\l/dr=(NB)^{1/2}\sim (Y_--Y)^{1/(2+Y_+)}\sim r^{1/(1+Y_+)}.
\eeq
The affine parameter distance to $S^\pm$ is therefore finite for all
$Y_+\in(-2,-1)$, corresponding to all $c_1\in(0,2)$.

Note that since the minimal 2-sphere is not hidden by a horizon, a
spherical congruence of null rays will converge towards the minimal
sphere and exit the other side with a positive expansion. The
Raychaudhuri equation shows that this can happen only if
$R_{ab}k^ak^b<0$ somewhere along the congruence, so we infer that
the aether stress tensor must violate the null energy condition in
this solution. We computed the curvature for this solution and found
that $G_{tt}=-\nu Y^2/Br^2$, so the energy density of the aether
($\propto G_{tt}$) is negative {\it everywhere} . The solution
nevertheless has positive total mass, which may at first seem to be
inconsistent but it is not. The total mass of an asymptotically flat
spacetime is given by a surface integral at spatial infinity, which
for stationary spacetimes is proportional to the volume integral
$\int_\Sigma R_{ab}n^a \xi^b dV$, plus a surface term if there is an
inner boundary~\cite{WaldBook}. Since $R_{ab}\propto T_{ab}-(1/2) T
g_{ab}$, it is not the energy density that figures in the total
energy but rather $R_{tt}$. Quite surprisingly, it turns out that
$R_{tt}$ vanishes everywhere in the static aether solution. (The
only nonzero component of the Ricci tensor is $R_{rr}$.) Hence the
energy integrand vanishes identically, as in Schwarzschild
spacetime. This does not mean the total energy vanishes however,
since there is a contribution from the inner boundary. In
Schwarzschild that inner boundary may be pushed off to the
asymptotic region on the other side of the Einstein-Rosen bridge,
but the static aether solution is singular on the other side of the
throat. One can think of the mass as determined by a boundary
condition at this singularity.

Let us briefly consider solutions for $c_1$ in the range
$3/2<c_1<2$, corresponding to $3/16<\n<1/4$. In this case the
distance to the internal infinite radius sphere is infinite, and all
the algebraic and differential invariants that we checked (including
$R$, $R_{ab}R^{ab}$, $R_{abcd}R^{abcd}$, $R_{ab}u^au^b$, and
$(\nabla_a u^b)(\nabla_b u^a)$) are asymptotically zero. However,
the curvature component $R_{ab}k^a k^b$ blows up asymptotically,
when $k^a$ is the tangent to an affinely parameterized radial null
geodesic approaching the internal infinity. The invariant $u^a k_a$
blows up as $N^{-1/2}$, since $N^{1/2}u^a$ is the Killing vector
$\xi^a$ and $k^a\xi_a$ is conserved along the geodesic. This
suggests that the above-mentioned invariants vanish because the
tensor structure of the curvature, the aether, and all derivatives
is determined by a single null vector pointing in the future radial
null direction opposite to $k^a$, i.e. pointing away from the
internal area-infinity.

Returning now to the generic solution for $0<c_1<1$, we examine more
closely the behavior at the throat and at the internal infinity.
Since the spherical radius $r$ is not a good coordinate at the
minimal area sphere, nor at the internal area-infinity, we adopt
instead the proper length coordinate $l$, in terms of which the line
element takes the form
\beq ds^2 = N(l) dt^2 - dl^2 - r(l)^2 d\Omega^2
\label{proplmetric}\eeq
To get an idea of how the throat geometry depends on $c_1$, we fix
the mass $M$ of the solution and plot in Fig.~\ref{Rcurves} the
numerically computed function $r(l)$ for several different values
of $c_1$. There is a discontinuity at $c_1=0$ where the solution
abruptly changes from a singular flare-out in finite proper
distance to a perfectly regular Einstein-Rosen bridge. The
singularity approaches the throat as $c_1\rightarrow0$, but in the
same limit the curvature becomes finite and the other half of the
bridge suddenly appears.
\begin{figure}
 \includegraphics[angle=-90,width=15cm]{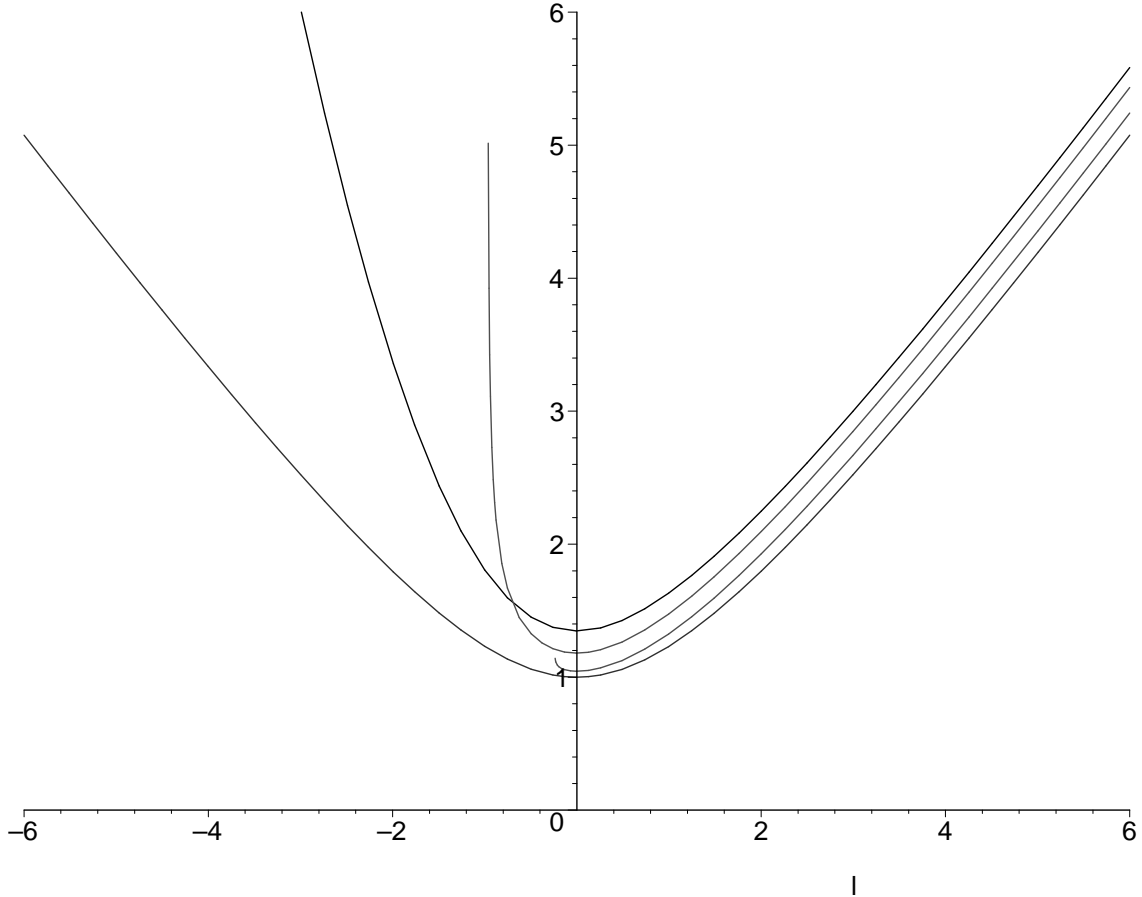}\\
\caption{\label{Rcurves} Plot of area radius $r$  vs. proper
length $l$ for fixed mass $M$, in units with $2M=1$, for $c_1 =$
0, 0.1, 0.7, and 1.9. In the GR case $c_1=0$ this is the
Einstein-Rosen bridge. For $c_1=0.1$ the radius flares out to
infinity so quickly that the code used to make the plot halted at
small radius. With increasing $c_1$ the throat widens, the
flare-out inside is slower, and the proper length to the curvature
singularity increases, becoming infinite for $c_1\ge3/2$.}
\end{figure}

To more fully compare the Schwarzschild and aether solutions we
plot together in Fig.~\ref{wormhole} the radius $r(l)$ and the
norm of the Killing vector $\sqrt{N(l)}$ for the two solutions
with the same value of the total mass $M$.
\begin{figure}
\includegraphics[angle=-90,width=15cm]{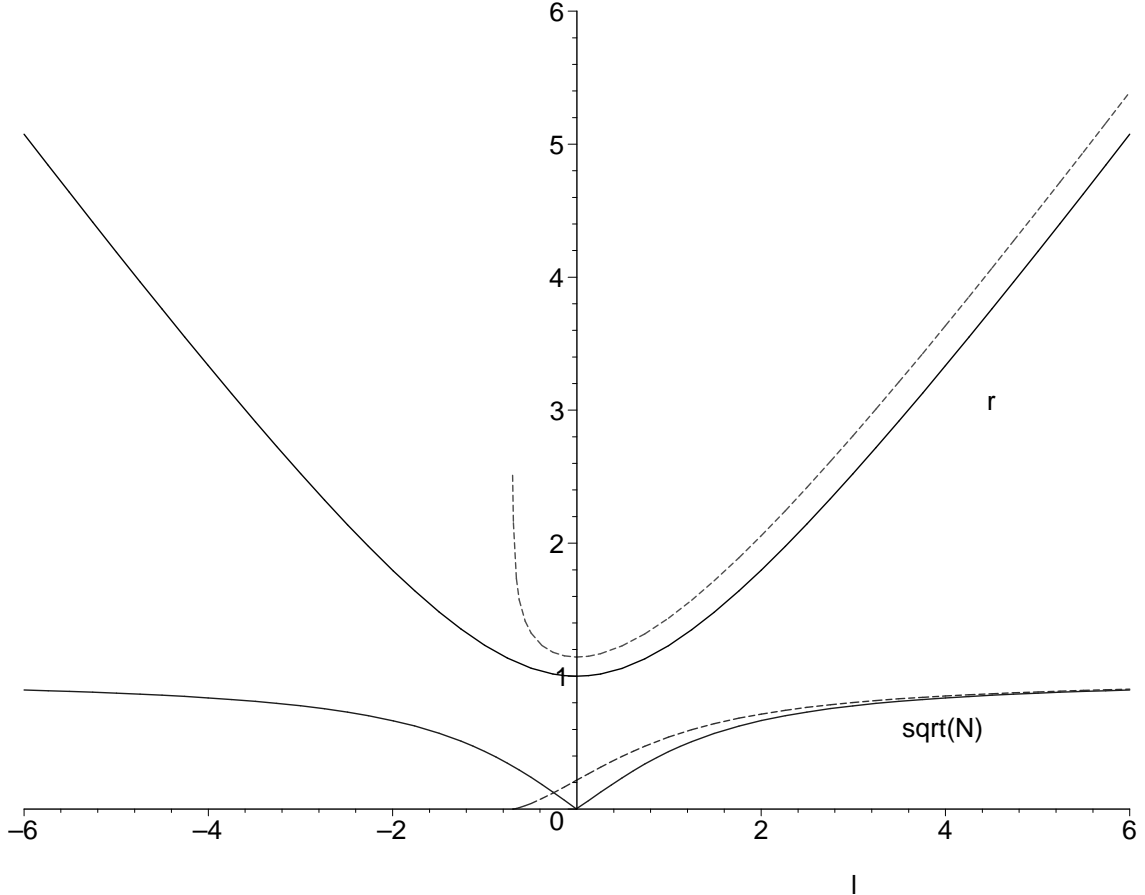}\\
\caption{\label{wormhole} Plot of $r(l)$ and the norm of the
Killing vector $\sqrt{N(l)}$ for GR and for $c_1=0.5$, for the
solution with the same value of the total mass $M$, in units with
$2M=1$. In GR $N$ vanishes at the bifurcation sphere at the center
of the Einstein-Rosen bridge. In the ae-theory solution the
Killing vector remains timelike at the throat, but at the internal
$r=\infty$ curvature singularity both the norm and its slope
vanish, indicating the presence of a singular extremal Killing
horizon.}
\end{figure}
At the internal singularity the norm of the Killing vector goes to
zero, and the Killing vector is tangent to the constant $r$
surfaces, so the singularity is a ``would-be" Killing horizon. The
surface gravity of the horizon is given by $d\sqrt{N}/dl$ at the
horizon. The behavior is easily found using the previous formulae:
\beq
d\sqrt{N}/dl= (1/2)(N/B)^{1/2}(Y/r)
\propto (Y-Y_-)^{2(1+Y_+)/(2+Y_+)}.
\eeq
The exponent is positive, so the derivative of the norm vanishes
as $Y\rightarrow Y_-$. The surface gravity is therefore zero, so
the would-be Killing horizon is extremal.

\subsubsection{Charged dust interpretation}
\label{dust}

We argued at the beginning of this section that the $c_2$ and
$c_3$ terms in the action do not contribute to the field equations
in the case of static, spherical symmetry and static aether, and
the $c_4$ term can be absorbed into a simultaneous shift of $c_1$
and $c_3$. This enabled us to reduce the general case to the one
with only $c_1$ non-zero. We can nevertheless choose to  include a
$c_3$ term in the action,  and by so doing re-express the content
of the field equations in an interesting way. In particular, if we
choose $c_3=-c_1$, then the $c_1$ and $c_3$ terms combine to make
$(c_1/2)F_{ab}F^{ab}$, where $F_{ab}= \nabla_a u_b-\nabla_b u_a$.
This is just the Maxwell Lagrangian for a vector potential $u_a$,
up to a constant factor. We have been treating the contravariant
vector $u^a$ as the independent field variable, but in this
Maxwell-like case it is natural to adopt instead the covariant
vector $u_a$ as independent. This change just amounts to an
invertible field redefinition, hence yields the same equations of
motion when the metric and aether field equations are taken
together. With this choice of field variable the theory with
$c_3=-c_1$ and $c_2=c_4=0$ looks quite similar to Maxwell theory,
the only difference being the constraint term
$\lambda(g^{ab}u_au_b-1)$ in the Lagrangian. It was shown
in~\cite{Jacobson:2000xp} that this is equivalent to the
Einstein-Maxwell-charged dust system, with a constant charge to
mass ratio fixed by $c_1$, and restricted to the sector in which
there exists a gauge choice for which the vector potential is
parallel to the dust 4-velocity. (This is a real restriction.)

It must be possible to interpret our strange static aether
wormhole solution as a charged dust solution, but it is at first
hard to see what could stabilize the dust unless it is extremally
charged, which corresponds to the case $c_1=2$. In fact it was
argued in Ref.~\cite{Jacobson:2000xp}, invoking prior
results~\cite{DeRay} for the charged dust problem, that there is
no static solution with static aether and general $c_1$. However,
our result here shows that argument cannot be correct. The
contradiction is resolved by the observation that the prior
results invoked in  Ref.~\cite{Jacobson:2000xp} apply only if the
dust mass density is positive, whereas in the solution at hand
this mass density turns out to be negative. (In Ref.~\cite{Bonnor}
the charged dust system with negative mass density and constant
charge to mass ratio was studied (among other cases) in static
axisymmetry, and it was shown that every harmonic function
determines a solution. Presumably among these solutions is the
static aether solution found here.) This is related to the
negative energy density that we already inferred above must be
present. With a negative mass density, the dust is gravitationally
repulsive, and though it has the same sign charge it is
electrically attractive. (In Newtonian terms, a force ${\bf F}$
produces an acceleration ${\bf F}/m$ which is opposite to ${\bf
F}$ when $m$ is negative.) Thus the gravitational and electric
forces exchange their usual roles. The fact that the dust does not
then just collapse on itself is perhaps due to the associated cost
in (positive) electric field energy when the dust is squeezed
together. Although static, the solution is not regular, since
there is an internal singularity.

\subsubsection{Solutions with $Y\in(Y_-,Y_+)$}
\label{nosoln}

So far we have discussed the solution for $Y$ in all ranges except
for $Y\in(Y_-,Y_+)$. In this range the metric function $B$ is
negative, so $\partial_r$ is timelike. Therefore, in order for the
metric signature to be Lorentzian, $\partial_t$ must be spacelike.
But in this static solution the aether is parallel to $\partial_t$
(cf. (\ref{statuform})), so cannot be timelike. Hence there is no
Lorentzian solution with timelike aether corresponding to this
range of $Y$.

As a mathematical curiosity, if we allow $N>0$, so there are two
timelike dimensions, there would still be a further restriction
for a real solution, since the ratio in (\ref{N}) is negative and
raised to the power $-Y_+/(2+Y_+)$. In order for $N$ to be real
and positive this power must be an even integer $m$, which implies
$\n=(1/4)(1-1/m^2)$. The ratio in (\ref{C/r}) is also negative,
and is raised to the different power $1/(2+Y_+)$, which can not
also be an integer since $Y_+$ is not an integer. However, the
integration constant $r_{\rm min}$ can be complex, thus balancing
the phase of the right hand side of (\ref{C/r}), and yielding a
real solution with signature (${+}{+}{-}{-}$).}

\section{Stars}
\label{stars}

In this section we investigate the static solutions to the
theory that, unlike the singular wormhole have a regular origin.
We first show that there are no such regular pure aether
solutions, and then turn to the case of fluid stars.

\subsection{Nonexistence of pure aether stars}
\label{solitons}

Spherically symmetric self-gravitating ``solitons" appear in a
number of field systems coupled to gravity. For example there are
boson star solutions in the Einstein-Klein-Gordon
system~\cite{Jetzer:1991jr} and Einstein-Yang-Mills theory
possesses the Bartnik-McKinnon
solutions~\cite{Bartnik:1988am,Volkov:1998cc}. It is therefore
natural to ask whether such``aether star" solutions might exist in
the vacuum Einstein-Aether theory. The static aether solution
studied in the previous section is the unique solution with static
aether (remember, this means aether aligned with the timelike
Killing vector), and does not have a regular origin. Thus the only
way a regular aether star might exist is if the aether has a
radial component. We now examine this possibility and show that it
cannot occur.

The analysis of Section \ref{Analysis} showed that local solutions
around a general $r=r_0$ are characterized by three free parameters
which may be taken to be $A'(r_0)$, $a(r_0)$ and $a'(r_0)$. If we
apply this result at the origin $r_0=0$ the parameter freedom is
restricted. Spherical symmetry implies that at the origin the radial
component of the aether vanishes, $b(0)=0$. The normalization
constraint (\ref{constr1}) therefore fixes $a(0) = e^{-A_0/2}$, and
the $r$-derivative of this constraint implies that $A'(0)$ and
$a'(0)$ are not independent, but rather are related by
$A'(0)=-2a'(0)/a(0)$. Thus there is a one parameter family of
solutions regular at the origin.

These solutions cannot be asymptotically flat, for the following
reason. The asymptotically flat boundary condition discussed in
Section \ref{Analysis} would require fixing the one free parameter,
leaving a unique solution. However, pure ae-theory is scale
invariant, so there must be at least a one parameter family of
solutions much like the Schwarzschild solutions of different mass in
GR. (By contrast, Einstein-Yang-Mills theory is {\it not} scale
invariant, and admits a discrete family of soliton and black hole
solutions.) We conclude that no regular aether stars exist. This
conclusion was verified empirically by integrating out from the
origin with different initial parameters, and attempting
unsuccessfully to tune to an asymptotically flat solution.

\subsection{Fluid Stars}
\label{fluidstars}

Although there are no regular  vacuum aether stars,  globally
regular solutions exist in the presence of a static, spherically
symmetric perfect fluid with no aether couplings. The fluid stress
tensor appearing in the metric field equation (\ref{AEE}) is
\beq T^M_{ab} = \bigl(\rho(r)+P(r)\bigr) v_a v_b - P(r)
g_{ab}\label{perffluid}\eeq
where $v^a = e^{-A/2}(\partial_t)^a$ is the fluid 4-velocity,
$\rho(r)$ it's mass density, and $P(r)$ it's pressure. The metric
field equation and the Bianchi identity together imply that the sum
of the aether and fluid energy-momentum tensors is divergenceless.
In addition, since the aether does not couple directly to the fluid,
its stress tensor is independently divergenceless when its field
equation and unit constraint are satisfied. Therefore the fluid
stress tensor is also independently divergenceless in any solution.
Thus, an appropriate system of equations for the aether plus fluid
case is the (i) metric field equation, (ii) aether field equation,
(iii) radial component of $\nabla^a T^M_{ab}=0$, which is the
hydrostatic equilibrium equation for the fluid
\beq P'+\half A'(\rho+P) = 0, \label{hydro} \eeq
and (iv) an equation of state $\rho=\rho(P)$.

The count of parameters that can be freely specified when
integrating out from a regular origin is the same as for the
vacuum aether solution, except for the addition of a central value
for the pressure $P_0=P(0)$. For a fixed central pressure, there
is just one parameter which can be tuned to obtain an
asymptotically flat solution. Now there is no contradiction with
scale invariance, since the central pressure sets the scale and
determines the total mass of the solution. That these
asymptotically flat star solutions have a static aether can be
inferred as follows. The field equations can be integrated out
from the origin with the aether assumed static. The pressure drops
to zero at some value $r=R$, the surface of the star, where the
static interior solution can be matched to the static vacuum
aether solution discussed in section \ref{static}. This solution
is asymptotically flat, so it must be the unique asymptotically
flat solution whose existence is indicated by the parameter count.

A possible worry about the preceding argument is whether the
matching to a vacuum solution at the surface of the star could be
non-unique. Provided  $a(r)$, $A(r)$, $B(r)$ and their derivatives
are continuous at $r=R$ one can simply continue the integration into
the vacuum region. We have not studied this behavior in detail, but
it seems that as long as this timelike surface is not a
characteristic surface of the ODE's corresponding to a spherical
wavefront of the spin-0 modes of the theory, the field equations
will imply the continuity condition.

To determine a star solution one can fix a central pressure and
numerically integrate the $tt$ and $rr$ metric field equations and
the hydrostatic equilibrium equation (\ref{hydro}), using the
equation of state, from the origin to the radius $R$ where the
pressure vanishes. There $A'(r)$ is continuous so one can use it to
match to the vacuum solution. The total mass $M$ can be read off
from (\ref{M}) together with (\ref{C/r}), using the definition
(\ref{A'}), $Y(R)=RA'(R)$. The area of the 2-spheres in such a star
solution is strictly increasing as $r$ increases from zero to the
surface of the star where $P$ vanishes. At that point $P'\le0$, so
according to (\ref{hydro}) $A'\ge0$ (assuming positive fluid energy
density $\rho$). Thus (\ref{A'}) implies $Y\ge0$, which means that
we always match to the static aether solution {\it outside} of the
minimal area 2-sphere. A ``throat" never occurs in such a star
solution.

\subsubsection{Constant density stars}

To get a sense of the nature of the static aether star solutions
we consider here the simplest example, stars with
constant energy density interior. Although this
does not closely describe realistic stars, it is adequate
for indicating the behavior of
maximum mass limits and the stability properties of
equilibrium configurations.
The discontinuity in the mass density at the surface
entails via the field equations a jump in $A''$,
but $A'$ remains continuous so can be used to match
to the vacuum solution as described above.

\begin{figure}
 \includegraphics[angle=-90,width=15cm]{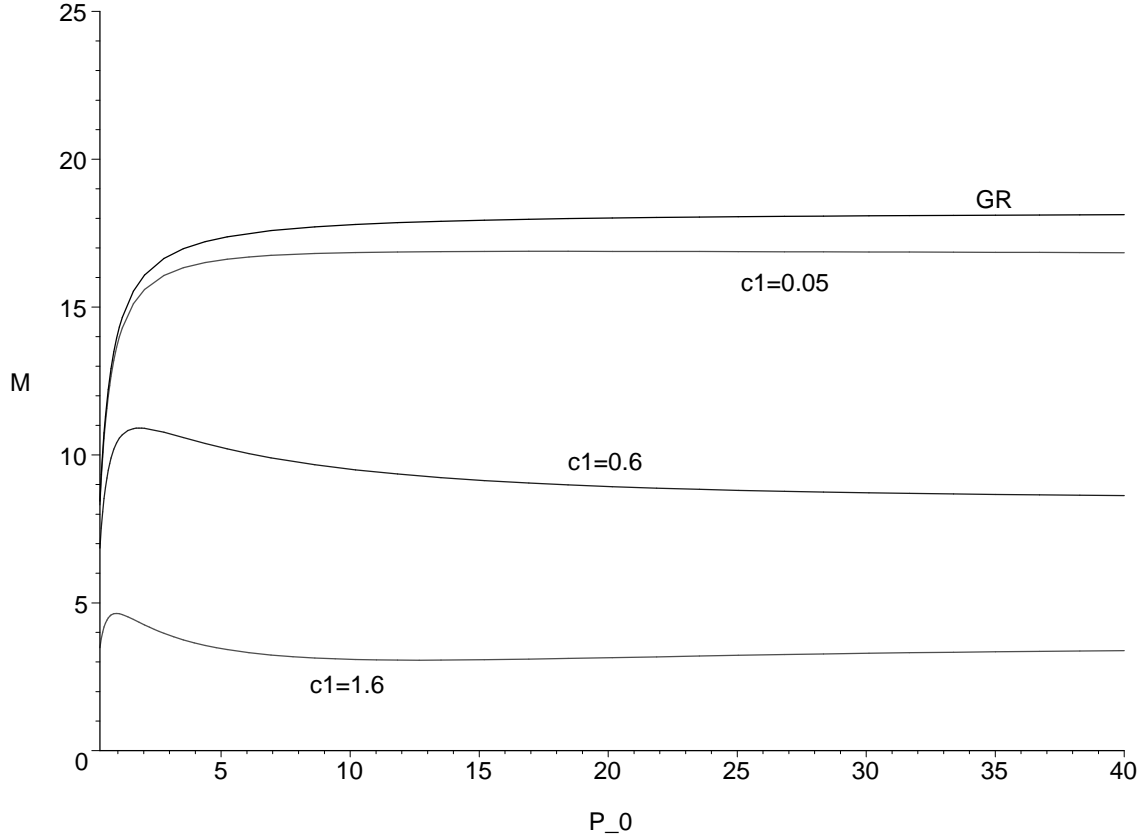}\\
\caption{\label{masses} Total mass vs.\ central pressure $P_0$
in a constant density star, in units with $\rho=1$ and $8\pi
G =1$, for several values of $c_1$.
The GR curve asymptotically approaches $M = (4\pi/3)
(24/9)^{3/2}$. As $c_1$ grows the maximum
mass decreases, and the curve develops a sharp local maximum
and a shallow local minimum. }
\end{figure}

Graphs of $M$ versus the central pressure $P_0$ for the equilibrium
configurations are displayed in Figure \ref{masses}.
In GR the mass asymptotes to a maximum value as the central pressure
goes to infinity. Physically, an infinite central pressure would be
required to maintain equilibrium for a greater mass. As $c_1$
increases in the Einstein-Aether case the maximum mass limit
decreases, and the mass curve develops a local maximum and a very
shallow local minimum that is only apparent for larger values of
$c_1$. For sufficiently large $c_1$ a second local maximum occurs.
(We have not attempted to determine the behavior at arbitrarily high
pressures and for $c_1$ approaching 2. Perhaps the series of maxima
and minima continues.)

\begin{figure}
 \includegraphics[angle=-90,width=15cm]{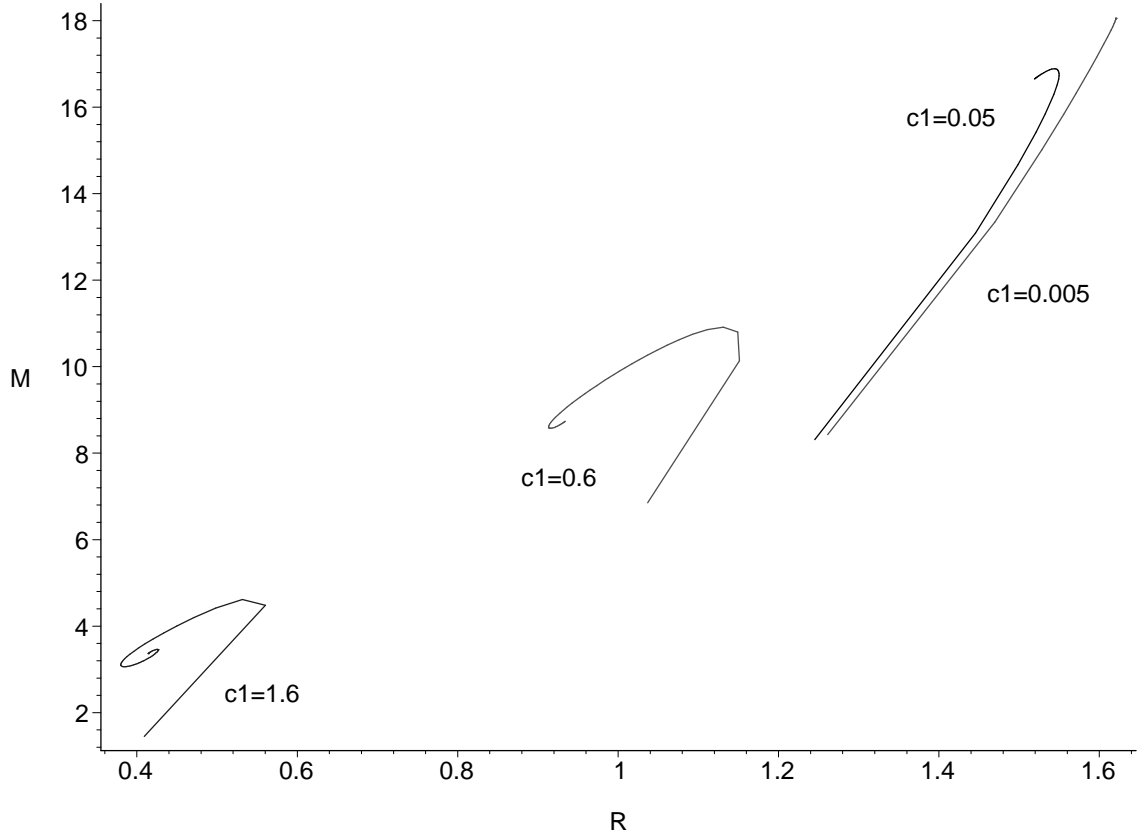}\\
\caption{\label{massvR} Total mass vs.\ $R$ for a constant density
star, in units with $\rho=1$ and $8\pi G =1$, for $P_0$ up to 300
and several values of $c_1$. For small pressures all the curves
increase uniformly. For $c_1=0.005$ the slope is nearly the same
as in GR. For $c_1=0.05$ a maximum occurs for a large central
pressure. By $c_1=0.6$, the first maximum occurs at much smaller
pressures, and there is also a minimum. For $c_1=1.6$ a second
maximum has appeared.}
\end{figure}

The presence of stationary points in the mass versus pressure curves
is an indication that the stability character of the equilibrium may
be changing~\cite{HartleBook}. The connection with linearized
stability arises as follows. If the squared frequency of a mode is
positive then the corresponding perturbation of the star is
oscillatory, while if it is negative the perturbation grows
exponentially in time. In the borderline case of zero frequency the
mode has zero energy, hence corresponds to a variation between two
static solutions with the same mass. Transitions between stability
and instability therefore occur at extrema of the mass versus $R$
plot shown in Figure \ref{massvR}, where a small displacement of $R$
does not change the mass to first order.

In the GR limit there is no critical point. The mass increases
monotonically with central pressure, as seen in Fig.~\ref{masses},
so there is no onset of instability. In ae-theory even flat space
is not necessarily stable. The conditions on the $c_i$ for which
all  linearized plane wave modes have positive squared frequency
were found in~\cite{Jacobson:2004ts}. For example in the pure
$c_1$ case they are $-2<c_1<1$. If we assume the values of $c_i$
are such that very small mass stars are stable, then instability
can only set in at a critical point of the mass function.  As
$c_1$ grows larger, the curves in Fig.~\ref{massvR} exhibit
extremal points corresponding to the local maxima and minima of
Fig.~\ref{masses}. At the maximum mass the lowest mode becomes
unstable. At the following local minimum another zero frequency
mode occurs, corresponding to the next mode becoming unstable. (It
cannot be the lowest mode becoming stable again, since $R$ is
increasing with increasing central pressure, implying the presence
of a node in the corresponding radial mode~\cite{HartleBook}.)
Therefore, beyond the maxima in shown in Figure \ref{massvR}
constant mass density stars in the Einstein-Aether theory are
unstable. For small values of $c_1$ the central pressure has to be
very large compared to the density for the star to reach the
instability, implying a violation of the dominant energy
condition. For $c_1=1$ the pressure at the onset of instability is
about 1.28 times the density.

\section{Discussion}
\label{discussion}

In this paper we analyzed the Einstein-Aether theory assuming
stationary spherical symmetry. We determined the number of free
parameters in the corresponding solutions to the field equations,
and classified the asymptotically flat ones. The vacuum solution
with static aether, i.e. aether aligned with the timelike Killing
vector, was found analytically up to inversion of a transcendental
equation. It has a wormhole-like structure, with a minimal 2-sphere
and a singular internal area-infinity which lies at finite affine
parameter along a radial null geodesic and finite proper distance if
$c_1<3/2$. The static fluid star solutions were classified and
studied numerically for the case of constant density, and the
maximum mass and stability properties were determined.

Several directions for further work are suggested by these
results. Foremost from an observational point of view might be to
determine the maximum mass limits for a realistic neutron star
equation of state. For astrophysical applications it would also be
necessary to determine the structure of rotating solutions.
Another open issue is the condition on the $c_i$ required by
stability of small mass stars. A related question is whether the
pure static aether wormhole solution is stable.

The solutions studied here may be of some help on the unresolved
question of energy positivity in ae-theory. It is known for what
ranges of the coefficients $c_i$ the energy of linearized solutions
is positive~\cite{Lim:2004js,Eling:2005zq}, and it is known that for
the Maxwell-like special case ($c_3=-c_1$, $c_2=c_4=0$) nonsingular
negative energy initial data
exist~\cite{Clayton:2001vy,Eling:2004dk}. The positive mass static
aether solutions on the other hand have positive energy despite
having everywhere negative energy density and an interior
singularity, as explained in section \ref{genericc1}.  It would be
interesting to see whether a positive energy result can be
established for nonsingular static, spherically symmetric star
solutions.

Finally, we found that there is a two parameter family of static,
asymptotically flat vacuum solutions. Stellar exteriors for
different masses form a one-parameter family. The other parameter
pertains to the radial tipping of the aether away from the Killing
vector. In a black hole solution the aether must tip, since it
cannot remain timelike and be aligned with the null Killing vector
on the horizon. In the companion to this paper~\cite{Eling:2006ec} we
determine the structure of black hole solutions in Einstein-Aether
theory.

\section*{Acknowledgements}
This research was supported in part by the NSF under grant
PHY-0300710.

\end{document}